\documentclass[pdflatex, referee, sn-mathphys]{sn-jnl}

\usepackage{multirow}
\usepackage{booktabs}
\usepackage{graphicx}
\usepackage{amsmath}
\usepackage{hyperref}

\jyear{2023}%

\begin{document}

\title{On the geometric phases in entangled states}

\author*[1]{\fnm{M. E.} \sur{Tunalioglu}}\email{tunalioglu@ankara.edu.tr}

\author[1]{\fnm{H. O.} \sur{Cildiroglu}}

\author[1]{\fnm{A. U.} \sur{Yilmazer}}

\affil[1]{\orgdiv{Department of Physics Engineering}, \orgname{Ankara University}, \orgaddress{\street{D\"{o}gol St.}, \city{Ankara}, \postcode{06100}, \state{Ankara}, \country{T\"{u}rkiye}}}

\abstract{Correlation relations for the spin measurements on a pair of entangled particles scattered by the two separate arms of interferometers in hybrid setups of different types are investigated. Concurrence, entanglement of formation, quantum fidelity, Bures distance are used to clarify how the geometric phase affects the initial bipartite state. This affect causes a quantum interference due to the movement of charged particles in regions where electromagnetic fields are not present. We shown that in some cases the geometric phase information is carried over to the final bipartite entangled state.}


\pacs[Physics and Astronomy Classification]{03.65.Ta, 03.65.Ud, 03.65.Vf}


\maketitle

\section{Introduction}\label{sec1}

In classical and quantum physics, the geometric phase is a unifying and innovative concept, and it appears as an additional phase factor due to the geometric and topological properties of the Hamiltonian parameter space for various driven systems. The geometric phase was first studied by Ehrenberg \cite{Ehrenberg_1949}, Kato \cite{doi:10.1143/JPSJ.5.435}, Pancharatnam \cite{pancharatnam1956generalized}, Longuet-Higgins et al. \cite{longuet1958studies}, Aharonov and Bohm \cite{aharonov1959significance} and later generalized by Berry in 1984 \cite{berry1984quantal}. Geometric and topological phases play a considerable role in physics and a vast literature exist today covering many plausible applications in different fields \cite{aharonov1984topological,PhysRevA.47.3424,PhysRevLett.72.5,PhysRevLett.83.2486,sponar2010geometric,sponar2010new,lepoutre2012he,Werner2012,gillot2013measurement,cohen2019geometric}. On the other hand, entanglement is another striking aspect of quantum mechanics and its dates back to the article published by Einstein et.al. \cite{einstein1935can} and its deep meaning was exhibited later by Bell \cite{bell1964einstein}.

The study of topological phases for entangled states has gained notable interest recently and it has been considered as a useful resource to analyze the quantum information processing. From this perspective, in this work, quantum correlations are briefly discussed for entangled particles in hybrid Aharonov-Bohm (AB), Aharonov-Casher (AC), He-McKellar-Wilkens (HMW), Dual Aharonov-Bohm (DAB) type Einstein, Podolsky and Rosen (EPR) setups. In this study, Clauser-Horne-Shimony-Holt (CHSH) inequality \cite{clauser1969proposed} for certain choices of spin measurements is discussed in a hybrid AC-EPR setup and then various entanglement measures such as concurrence ($\mathcal{C}$), entanglement of formation ($EoF$) are examined \cite{bennett1996mixed, wootters1998entanglement, WERNER2006233, 10.5555/2011706.2011707} in connection with the quantum states acquiring geometric and topological phases. At the same time, in order to better understand the possible role of geometric phase as an indicator of the entanglement concepts such as quantum fidelity ($\mathcal{F}$) and Bures distance ($\mathcal{D}_B$) are utilized.

\section{An Examplary Model: Entangled State in a Hybrid AC-EPR Setup}\label{sec2}
\begin{figure}[H]
\includegraphics[width=\textwidth]{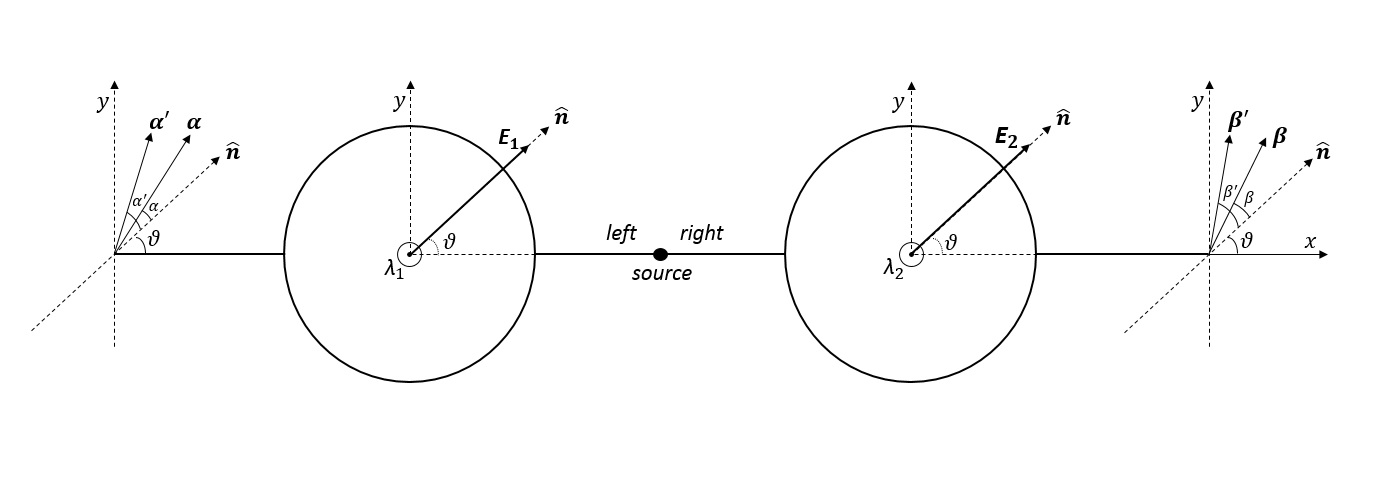}
\caption{A hybrid AC-EPR setup. To investigate the result of the measurement of the spin components of the particles going left and right, one can deﬁne projection operators along arbitrary vectors $\alpha$, $\alpha'$, $\beta$ and $\beta'$. Along these vectors the expectation values of joint measurements necessary for the CHSH inequality are calculated.}
\label{AC-EPR}
\end{figure}

In order to investigate the quantum spin correlations related to the Aharonov-Casher effect performed with an entangled spin pairs a hybrid AC-EPR experimental setup with two electric charge lines ($\lambda_1,\lambda_2$) provides a useful and practical configuration (see Fig. \ref{AC-EPR}). Consider a pair of entangled spin-$\frac{1}{2}$ neutral particles produced in a spin singlet state by the source at the center:
\begin{equation}
\label{eqn:spin_singlet}
    \vert\psi(t)\rangle = \frac{1}{\sqrt{2}}[\vert \! \uparrow \rangle_L\otimes \vert \! \downarrow \rangle_R - \vert \! \downarrow \rangle_L\otimes \vert \! \uparrow \rangle_R ].
\end{equation}
Thus we are considering a \textit{gedanken} experiment with two magnetic dipoles (neutrons) from a single source, one to the right, one to the left, and are under the influence of two electric charge lines.

Our goal is to discuss the possible results of the spin measurements in various directions and to find the correlation functions associated with the joint probabilities; and furthermore we will investigate the CHSH violation, calculate the concurrence ($\mathcal{C}$), entanglement of formation ($EoF$) and quantum fidelity ($\mathcal{F}$), Bures distance ($\mathcal{D}_B$) step by step.

The initial singlet state in \eqref{eqn:spin_singlet} will gain an AC phase as follows,
\begin{equation}
\label{eqn:ac_phase1}
    \vert \psi(t') \rangle = \frac{1}{\sqrt{2}} [e^{-i\mu\lambda_1} \vert \! \uparrow \rangle_L\otimes e^{i\mu\lambda_2} \vert \! \downarrow \rangle_R \\ - e^{i\mu\lambda_1} \vert \! \downarrow \rangle_L \otimes e^{-i\mu\lambda_2} \vert \! \uparrow \rangle_R].
\end{equation}
\noindent This equation can be written more simply,
\begin{equation}
    \vert\psi(t')\rangle = \frac{1}{\sqrt{2}}[e^{-i\mu(\lambda_1-\lambda_2)} \vert \! \uparrow \rangle_L \vert \! \downarrow \rangle_R \\ - e^{i\mu(\lambda_1-\lambda_2)} \vert \! \downarrow \rangle_L \vert \! \uparrow \rangle_R] \label{eqn:ac_phase2}.
\end{equation}
\noindent If one defines $\lambda_1 - \lambda_2 = \lambda_E$, then above equation becomes,
\begin{equation}
\label{eqn:ac_phase3}
    \vert\psi(t')\rangle = \frac{1}{\sqrt{2}}e^{-i\mu\lambda_E}[ \vert \! \uparrow \rangle_L \vert \! \downarrow \rangle_R - e^{2i\mu\lambda_E} \vert \! \downarrow \rangle_L \vert \! \uparrow \rangle_R].
\end{equation}
\noindent The total phase factor can obviously be removed as it will not be included in the expectation value calculation, and the final spin wave function will be the following,
\begin{equation}
\label{eqn:ac_phase4}
    \vert\psi(t')\rangle = \frac{1}{\sqrt{2}}[
\vert \! \uparrow \rangle_L \vert \! \downarrow \rangle_R - e^{2i\mu\lambda_E} \vert \! \downarrow \rangle_L \vert \! \uparrow \rangle_R].
\end{equation}

The quantum correlations can be examined by measuring the spin components of the particles moving to the right and left, along the directions as in the usual Bell tests. The correlation function ($S$) for the CHSH inequality associated with the angles ($\alpha, \beta, \alpha', \beta'$) for the spin measurements shown in \autoref{AC-EPR} can be found to be as:
\begin{equation}
\begin{split}
\label{eqn:s_func1}
    S(\alpha, \beta, \alpha', \beta') = \vert E(\alpha,\beta)-E(\alpha,\beta')\vert+\vert E(\alpha',\beta)+E(\alpha',\beta')\vert \\ = \vert-\cos\alpha\cos\beta - \sin\alpha\sin\beta\cos(2\mu\lambda_E) \\ + \cos\alpha\cos\alpha' + \sin\alpha\sin\alpha'\cos(2\mu\lambda_E)\vert \\ + \vert-\cos\alpha'\cos\beta' -  \sin\alpha'\sin\beta'\cos(2\mu\lambda_E) \\ + \cos\beta\cos\beta' + \sin\beta\sin\beta'\cos(2\mu\lambda_E)\vert \leq 2.
\end{split}
\end{equation}
\noindent Here $E\boldsymbol{(\alpha, \beta)}$'s are expectation values of the joint spin measurements defined as, $E\boldsymbol{(\alpha, \beta)} = \vert\langle\psi(t') [P^{l}_{+}\boldsymbol{(\alpha)}-P^{l}_{-}\boldsymbol{(\alpha)]}\otimes[P^{r}_{+}\boldsymbol{(\beta)}-P^{r}_{-}\boldsymbol{(\beta)]} \vert\psi(t')\rangle\vert$ and $P^{l}_{\pm}\boldsymbol{(\alpha)}$ being the spin projection operators along the angles $\boldsymbol{{\pm}(\alpha)}$, etc. \cite{CILDIROGLU2021127753}.

It is easily seen that the CHSH inequality is violated by quantum mechanics, if the appropriate angles are chosen for the $S$ function. At this point, one notes that Bell-type inequalities have a generalized upper violation limit called Tsirelson limit $(\vert\langle E \rangle\vert_{QM correlations} \leq 2\sqrt{2})$ \cite{cirel1980quantum}, and for appropriate choices of relevant angles below, one obtains:

\begin{equation}
\label{eqn:s_func3}
    S(0, \frac{\pi}{4}, \frac{3\pi}{4}, \frac{\pi}{2}) = \sqrt{2} + \sqrt{2}\vert\cos(2\mu\lambda_E)\vert.
\end{equation}
This result reveals the clear dependence of the CHSH correlation function $S$ on the geometric AC phase $2\mu\lambda_E$.

\section{Related Entanglement Measures}\label{sec3}
Now in order to discuss the entanglement measures in connection with the Bell-type correlations for the entangled states we continue to examine the above mentioned hybrid AC-EPR setup with two electric charge lines. Firstly, let us calculate concurrence. One needs to write a more suitable version of the wave function, namely $\vert\widetilde{\psi}(t')\rangle$, according to the definition given in \cite{wootters1998entanglement},
\begin{equation}
\label{eqn:concur_wave}
    \vert\widetilde{\psi}(t') \rangle = \sigma_y \otimes \sigma_y \vert\psi^*\rangle = \frac{1}{\sqrt{2}}[(\vert \! \downarrow \uparrow \rangle - e^{-2i\mu\lambda_E}\vert \! \uparrow \downarrow \rangle)].
\end{equation}
\noindent Pauli spin matrices here will spin flip the complex conjugate of the wave function, thus the concurrence becomes,
\begin{equation}
\label{eqn:concur1}
    \mathcal{C}(\vert\psi(t')\rangle) = \left\vert\langle \psi(t')\vert\widetilde{\psi}(t')\rangle \right\vert = \\ \frac{1}{2}\left\vert(\langle \uparrow \downarrow \! \vert - e^{-2i\mu\lambda_E}\langle \downarrow \uparrow \! \vert)(\vert \! \downarrow \uparrow \rangle - e^{-2i\mu\lambda_E} \vert \! \uparrow \downarrow \rangle)\right\vert.
\end{equation}
\noindent If we arrange the above result, and performs simple inner products the concurrence is found as
\begin{equation}
\label{eqn:concur3}
    \mathcal{C}(\vert\psi(t')\rangle) = \left\vert -e^{-2i\mu\lambda_E} \right\vert = 1.
\end{equation}

Thus, the entanglement of formation, $EoF$, is defined via the concurrence value and the function $h(x) = -xlog_2x -(1-x)log_2(1-x)$ (see ref. \cite{wootters1998entanglement}). If we use the above result in the $EoF$ calculation \eqref{eqn:concur3} we get
\begin{equation}
\label{eqn:eoafc}
    EoF =h\left ( \frac{1+\sqrt{1-\mathcal{C}^2(\rho)}}{2} \right ) = h \left ( \frac{1}{2} \right ) = 1.
\end{equation}

In the case when both states $(\vert\psi(t)\rangle$ and $\vert\psi(t')\rangle)$ are pure states then quantum fidelity can be calculated as an overlap of states,
\begin{equation}
\label{eqn:fidelity_ac1}
    \mathcal{F} = \vert\langle\psi(t)\vert\psi(t')\vert = \\ \frac{1}{2} \big\vert \langle \uparrow\downarrow \!\vert\! \uparrow\downarrow\rangle - e^{2i\mu\lambda_E}\langle \uparrow\downarrow \!\vert\! \downarrow\uparrow \rangle \\ - \langle \downarrow\uparrow \!\vert\! \uparrow\downarrow \rangle + e^{2i\mu\lambda_E} \langle \downarrow\uparrow \!\vert\! \downarrow\uparrow  \rangle \big\vert.
\end{equation}
\noindent Thus quantum fidelity for a process of hybrid AC-EPR scattering with spin entangled initial state turns out to be
\begin{equation}
\label{eqn:fidelity_ac2}
    \mathcal{F} = \frac{1}{2}\big\vert1+e^{2i\mu\lambda_E}\big\vert = \vert\cos(\mu\lambda_E)\vert.
\end{equation}
Furthermore, the Bures distance is given as $\mathcal{D}_B=\sqrt{2(1-\mathcal{F})}$ (see ref.\cite{yuan2017quantum}), so $\mathcal{D}_B$ between initial and final entangled states in a hybrid AC-EPR setup can be found as
\begin{equation}
\label{eqn:bures_dist}
    \mathcal{D}_B(AC)=\sqrt{2-2\vert\cos(\mu\lambda_E)\vert}.
\end{equation}

The state in \eqref{eqn:spin_singlet} is a pure entangled state of two qubits, so by Gisin's theorem \cite{GISIN1991201}, it violates the CHSH inequality and our result in \eqref{eqn:s_func3} confirms that prediction. Also by the equations \eqref{eqn:concur3}, \eqref{eqn:eoafc} and \eqref{eqn:fidelity_ac2}; one observes that the geometric phase has no effects on the concurrence and entanglement of formation, but on the contrary it has an explicit appearance in the quantum fidelity expression and the latter ranges between $0\leq \mathcal{F} \leq 1$. Also $\vert\psi(t)\rangle$ and $\vert\psi(t')\rangle$ which are connected via an AC phase as in \eqref{eqn:ac_phase4} and the non-zero Bures distance between them is given by the equation \eqref{eqn:bures_dist}.

In this context, AB, HMW, DAB type experimental hybrid setups can be designed similar to the hybrid AC-EPR setup and analogous calculations can be repeated for them. Thus, the summarized results, including a generalized version of the geometric Berry phase, can be obtained as written in Table \ref{table:table1}.

\begin{sidewaystable}
\sidewaystablefn%
\begin{center}
\begin{minipage}{\textheight}
\caption{Summary of Aharonov-Bohm (AB), Aharonov-Casher (AC), He-McKellar-Wilkens (HMW), Berry and Dual Aharonov-Bohm (DAB) phases gained by entangled states and the values of Concurrence ($\mathcal{C}$), Entanglement of Formation (\textit{EoF}), Quantum Fidelity ($\mathcal{F}$), Bures Distance ($\mathcal{D}_B$). Here $\mu$ is magnetic dipole, $\lambda_E$ is electric charge density, $d$ is electric dipole, $\lambda_B$ is magnetic charge density, $\gamma$ is Berry phase and $g$ is magnetic charge.}
\label{table:table1}
\begin{tabular}{@{}ccccccc@{}}
\toprule
\textbf{Phase} &
  \textbf{\begin{tabular}[c]{@{}c@{}}$\mathbf\vert{\psi(t)\rangle^{1,3}}$\end{tabular}} &
  \textbf{\begin{tabular}[c]{@{}c@{}}$\mathbf\vert{\psi(t')\rangle^{2}}$ \end{tabular}} &
  \textbf{$\mathcal{C}$} &
  \textbf{\textit{EoF}} &
  \textbf{$\mathcal{F}$} &
  \textbf{$\mathcal{D}_B$} \\ \midrule
\textbf{AB} &
  \multirow{5}{*}{\begin{tabular}[c]{@{}c@{}}\\ \\ \\ $ \frac{1}{\sqrt{2}}[ \vert \! \uparrow\rangle_L\vert \! \downarrow\rangle_R - \vert \! \downarrow\rangle_L\vert \! \uparrow\rangle_R]$\end{tabular}} &
  \begin{tabular}[c]{@{}c@{}}$\frac{1}{\sqrt{2}}e^{-i\phi_B}[\vert \! \uparrow \rangle_L \vert \! \downarrow \rangle_R - \vert \! \downarrow \rangle_L \vert \! \uparrow \rangle_R]$\end{tabular} &
  1 &
  1 &
  1 &
  0 \\ \cmidrule(r){1-1} \cmidrule(l){3-7} 
\textbf{AC} &
   &
  \begin{tabular}[c]{@{}c@{}}$\frac{1}{\sqrt{2}}[\vert \! \uparrow \rangle_L \vert \! \downarrow \rangle_R - e^{2i\mu\lambda_E} \vert \! \downarrow \rangle_L \vert \! \uparrow \rangle_R]$\end{tabular} &
  1 &
  1 &
  $\vert\cos(\mu\lambda_E)\vert$ &
  $\sqrt{2-2\vert\cos(\mu\lambda_E)\vert}$ \\ \cmidrule(r){1-1} \cmidrule(l){3-7} 
\textbf{HMW} &
   &
  \begin{tabular}[c]{@{}c@{}}$\frac{1}{\sqrt{2}}[\vert \! \uparrow \rangle_L \vert \! \downarrow \rangle_R - e^{2id\lambda_B} \vert \! \downarrow \rangle_L \vert \! \uparrow \rangle_R]$\end{tabular} &
  1 &
  1 &
  $\vert\cos(d\lambda_B)\vert$ &
  $\sqrt{2-2\vert\cos(d\lambda_B)\vert}$ \\ \cmidrule(r){1-1} \cmidrule(l){3-7} 
\textbf{Berry} &
   &
  \begin{tabular}[c]{@{}c@{}}$\frac{1}{\sqrt{2}}[\vert \! \uparrow \rangle_L \vert \! \downarrow \rangle_R - e^{2i\gamma} \vert \! \downarrow \rangle_L \vert \! \uparrow \rangle_R]$\end{tabular} &
  1 &
  1 &
  $\vert\cos(\gamma)\vert$ &
  $\sqrt{2-2\vert\cos(\gamma)\vert}$ \\ \cmidrule(r){1-1} \cmidrule(l){3-7} 
\textbf{DAB} &
   &
  \begin{tabular}[c]{@{}c@{}}$\frac{1}{\sqrt{2}}e^{-ig\phi_E}[\vert \! \uparrow \rangle_L \vert \! \downarrow \rangle_R - \vert \! \downarrow \rangle_L \vert \! \uparrow \rangle_R]$\end{tabular} &
  1 &
  1 &
  1 &
  0 \\ \bottomrule
\end{tabular}%
\footnotetext[1]{Initial State Vector at $t$}
\footnotetext[2]{Final State Vector at $t'$}
\footnotetext[3]{The entangled state at time $t$ for all five cases}
\end{minipage}
\end{center}
\end{sidewaystable}

Quantum computers are analog machines, which are expected to become a part of our lives in the near future. Although the codes to be run in the quantum computer are applied on various quantum gates, they are actually represented by real variables. Applying a quantum gate to a qubit means subjecting it to a wave from an indiscriminate wave generator. All properties of this wave, such as amplitude and frequency, are real-valued variables. Errors may occur from time to time in these variables. These errors are different from those caused by the interaction of the qubit with the environment. Even if a required quantum gate is prepared and applied perfectly, however exactly the desired quantum state may not be achieved.

In fact, this is where quantum fidelity comes into play. Fidelity is a link between regular digital needs and quantum hardware. A quantum state will be less entangled the closer it is to the set of separable states, or more entangled the farther it is \cite{liang2019quantum}. With fidelity calculations, one finds a way to track how well real circuits produced by a quantum computer fit the original state. Mathematically, it provides guidance on the degree between quantum states \cite{irtf-qirg-principles-10}, and there are several benefits for making such a comparison. As a result fidelity has become one of the most widely used quantities to measure the degree of similarity between quantum states.

\begin{figure}[H]
\includegraphics[width=\textwidth]{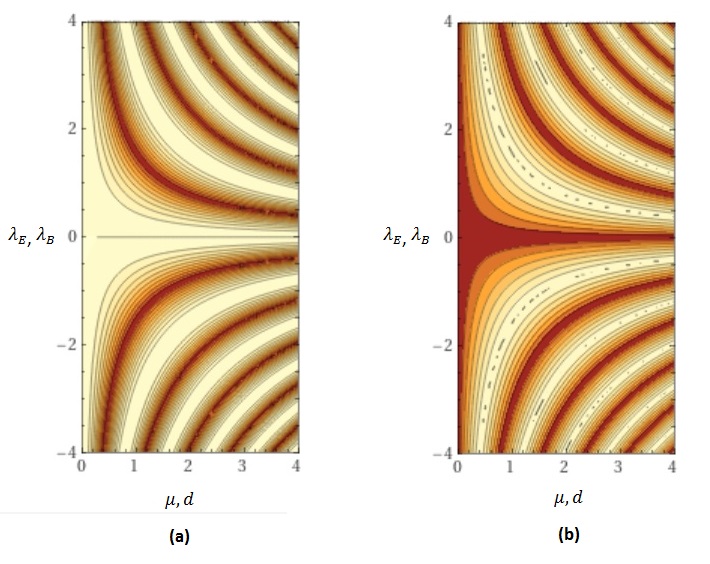}
\caption{Plots of (a) Fidelity and (b) Bures distance for AC-EPR and HMW-EPR Hybrid Setups, respectively. The white areas are the regions where the absolute value of the fidelity expression given for AC ($\vert\cos(\mu\lambda_E)\vert$) and HMW setups ($\vert\cos(d\lambda_B)\vert$) in Table \ref{Table1} attains unity, and in the dark areas are this absolute value becomes zero. Similarly, as to the Bures distance expression given for AC ($\sqrt{2-2\vert\cos(\mu\lambda_E)\vert}$) and HMW ($\sqrt{2-2\vert\cos(d\lambda_B)\vert}$), the region where the absolute value reaches unity is the dark areas, and conversely the white areas are the regions where the absolute value is zero. As a result, if the fidelity value is $1$, the Bures distance value is $0$ as expected. This tells us that the geometric phase information is preserved (observed) for the parameter values of the white areas but not preserved (seen) in the dark areas. In this plot exemplary $\mu$, $d$, $\lambda_E$ and $\lambda_B$ are the relevant physical variables  having certain values within some exemplary intervals chosen as ($\mu, d=[0,4]$ and $\lambda_E, \lambda_B=[-4,4]$) and each product $\mu\lambda_E$ and $d\lambda_B$ is an angle. It is to be noted that; since there is no geometric phase information in AB and DAB type hybrid experimental setups and neither in the Berry case, no figures are needed for them.}
\label{F-B}
\end{figure}

Besides, Bures distance describes the infinitesimal distance between the density matrices describing the quantum states, and this approach is a generalization of the Fisher distance. If it is limited to pure states only, it gives similar results as the Fubini–Study distance \cite{doi:10.1080/09500349414552171, HUBNER1992239, bures1969extension, HELSTROM1967101, facchi2010classical}. These distances can be used as a function of distance, metric. The Fisher metric is a Riemann metric that can be defined on a smooth statistical manifold. It is generally used to calculate the difference in information between measurements.

In this regard, statistical distances such as the Bures distance are determined by the size of the statistical fluctuations that occur in the measurements made to distinguish between the initial state and the evolved state. These statistical fluctuations raise the interesting possibility that quantum mechanics may be partially responsible for the Hilbert-space structure, so Bures distance is thought to be a link between statistics and geometry. By counting the number of intermediate states, the distance between states can be found \cite{wootters1981statistical}.

\section{Conclusion}\label{sec4}
Summary Table \ref{table:table1}  and \autoref{F-B} shows us that, AB and DAB phase information does not appear in the results of the measurements in the experimental setups to be made with entangled particles. However, the situation is different for AC, HMW and Berry phases. For these processes, geometric phase information appears in the joint spin measurements. Thus, it seems possible in the future to code an information via the geometric phase relevant in the process. It is understood that investigation of the entanglement properties by the geometric phase will provide a better understanding of the nature of entanglement and its role in quantum technologies \cite{sandhya2012geometric}.

Nevertheless, quantum logic gate operations can also be implemented with geometric phase. When a quantum system is in a cyclic evolution, it acquires a geometric phase determined by the path in which the system moves. Geometric phases are useful for combatting errors in quantum gate operations and can help quantum error correction codes reach below the error threshold. There are many studies examining how geometric phases can be used to better analyze entangled systems and to process the quantum information \cite{Vedral2003, Sjoqvist2015, Thomas2016}.

Since the geometric phase is not dependent on time and energy like the dynamic phase, but is connected to the closed path, it is not affected by changes such as noise distortions. Therefore, it can also be used for quantum logic gate applications \cite{Song2017, ji-li-na}. Obviously one of the main questions here is the entanglement content of the measured state, so after a physical process providing a geometric phase the different measures of entanglement are calculated for this purpose. 
 
Otherwise, as is well known interferometers of different types, such as Mach–Zehnder, Hanbury Brown and Twiss, etc. are effectively used to study counter intuitive predictions of quantum mechanics \cite{doi:10.1119/1.12387, Grangier_1986, SILVERMAN-BOOK}. In such works, one observes that the probability calculations based on the correlations between the number of particle counts at detectors have interesting similarities with the probabilities via spin measurements of the particles in certain directions as used in the present work.

\section*{Acknowledgments}
We thank Abdullah Vercin for helpful comments.

\section*{Declarations}
The authors did not receive support from any organization for the submitted work. The authors have no relevant financial or non-financial interests to disclose.

\bibliography{sn-bibliography}

\end{document}